\documentclass[aip,rsi,reprint]{revtex4-1}

\usepackage{graphicx}

\renewcommand{\d}{\mathrm{d}}

\newcommand{\rev}[1]{#1}

\begin{document}

%\preprint{}

\title{Calibration and energy measurement of \rev{optically levitated nanoparticle sensors}} %Title of paper
%\title{Displacement Calibration and Energy Measurement of Nonlinear Micro- and Nanomechanical Oscillators} %Title of paper



\author{Erik Hebestreit}
\homepage[]{http://www.photonics.ethz.ch}
%\thanks{}
\affiliation{Photonics Laboratory, ETH Z\"{u}rich, 8093 Z\"{u}rich, Switzerland}

\author{Martin Frimmer}
\affiliation{Photonics Laboratory, ETH Z\"{u}rich, 8093 Z\"{u}rich, Switzerland}

\author{Ren\'{e} Reimann}
\affiliation{Photonics Laboratory, ETH Z\"{u}rich, 8093 Z\"{u}rich, Switzerland}

\author{Christoph Dellago}
\affiliation{Faculty of Physics, University of Vienna, Boltzmanngasse 5, 1090 Wien, Austria}
\affiliation{Erwin Schr\"{o}dinger International Institute for Mathematics and Physics, University of Vienna, Boltzmanngasse 9, 1090 Wien, Austria}

\author{Francesco Ricci}
\affiliation{ICFO-Institut de Ciencies Fotoniques, The Barcelona Institute of Science and Technology, Castelldefels, Barcelona 08860, Spain}

\author{Lukas Novotny}
\affiliation{Photonics Laboratory, ETH Z\"{u}rich, 8093 Z\"{u}rich, Switzerland}


\date{\today}

\begin{abstract}

Optically levitated nanoparticles offer enormous potential for precision sensing.
%The absolute measurement performance of a levitated-particle sensor relies on the accuracy of the calibration relating the measured signal to an absolute displacement of the particle.
However, as for any other metrology device, the absolute measurement performance of a levitated-particle sensor is limited by the accuracy of the calibration relating the measured signal to an absolute displacement of the particle.
%Furthermore, the laws of thermodynamics put stringent requirements on the measurement time when interrogating a thermal state.
Here, we suggest and demonstrate calibration protocols for levitated-nanoparticle sensors. Our calibration procedures include the treatment of anharmonicities in the trapping potential, as well as a protocol using a harmonic driving force, which is applicable if the sensor is coupled to a heat bath of unknown temperature. \rev{Finally, using the calibration, we determine the center-of-mass temperature of an optically levitated particle in thermal equilibrium from its motion, and discuss the optimal measurement time required to determine said temperature.}

\end{abstract}


\maketitle %\maketitle must follow title, authors, abstract and \pacs

%\tableofcontents
\section{Introduction}

Electromagnetic radiation \rev{is amongst} the most powerful probes in our measurement toolbox. A prominent example is the interferometric detection of the minute distortions of space-time by gravitational waves.\cite{Abbott2016} Any probe disturbs the system it measures, and so does light by the forces it exerts on matter.\cite{Braginsky1995} These optical forces can deliberately be harnessed to manipulate microscopic objects. For example, dielectric particles can be trapped in a strongly focused laser beam.\cite{Ashkin1977,Ashkin1997,Neuman2004} Besides trapping the particle, the laser light is also scattered off the particle, providing an optical signal encoding the particle's position. \rev{Due to its small mass, such an optically trapped particle is an excellent model system} to study thermodynamic processes both in and out of equilibrium in the regime of strong fluctuations arising due to the coupling of the particle to a heat bath.\cite{Li2010,Berut2012,Gieseler2014,Martinez2016,Martinez2017}
%When trapped in a liquid or a dense gas, the fluctuating forces acting on the trapped particle originate from collisions with the molecules of the surrounding medium.
When trapped in a liquid or a dense gas, the fluctuating forces that originate from collisions with the molecules of the surrounding medium cause the trapped particle to \rev{undergo} overdamped Brownian motion.\cite{Berg2004}
At low gas pressures, an optically trapped particle \rev{behaves like} a strongly underdamped oscillator, whose sensitivity to perturbations close to its eigenfrequency is boosted by the quality factor of the mechanical resonance.\cite{Li2011,Gieseler2012,Kiesel2013,Moore2014,Millen2015,Vovrosh2017} In ultra-high vacuum, the dominant heat bath governing the thermodynamics of an optically trapped particle is the trapping laser and the quantum nature of the light field manifests itself as radiation pressure shot noise.\cite{Jain2016}

\rev{An optically levitated particle can ultimately be viewed as an extremely sensitive sensor.\cite{Moore2014,Ranjit2015,Ranjit2016}} A force acting on the particle gives rise to a displacement, which can be measured by observing the optical scattering signal. Importantly, for quantitative force measurements, we require a calibration that relates the detected optical signal, typically a voltage $V$, to the particle position $q$. For an optical signal that is linear in particle displacement, we therefore need to know the calibration constant $c_\mathrm{calib}$ fulfilling the relation $V = c_\mathrm{calib}\cdot q$.

\rev{The state-of-the-art calibration procedure for levitated nanoparticles} is to measure the particle's trajectory in thermal equilibrium and invoke equipartition of the potential energy amongst all degrees of freedom.\cite{Berg2004,Hauer2013} Three important points need to be kept in mind when using this method. First, the bath temperature needs to be known. Second, the trapping potential needs to be harmonic. Third, the measurement needs to be long enough to average out the fluctuations of the thermal state.
For typical experimental conditions, however, not all of these points are always fulfilled \rev{in particle-levitation experiments}. For example, at reduced gas pressures, the residual absorption of the trapped particle generates a significant heating of the particle's internal temperature, which \rev{leads to} an elevated (but a-priori unknown) effective bath temperature.\cite{Millen2014} Furthermore, a typical optical trapping potential has significant anharmonicities sampled by the particle at room temperature.\cite{Gieseler2013} Finally, the question \rev{of} a suitable measurement time to determine the particle's center-of-mass temperature in a thermal state has not been explicitly addressed by the community of levitated optomechanics.

In this paper, we suggest and experimentally demonstrate calibration procedures for sensors based on optically trapped nanoparticles. In particular, our discussion includes (1)~the calibration of sensors with non-harmonic trapping potentials \rev{and a known bath temperature}, (2)~the calibration for experimental conditions with unknown effective bath temperature, and (3)~a quantitative discussion of the measurement duration required to determine the center-of-mass temperature of a thermal state with a given confidence interval.

This paper is structured as follows: In Sec.~\ref{sec:Exp_setup} we introduce our experimental setup. Section~\ref{sec:calibration-of-the-detected-signal} describes how to calibrate a levitated nanoparticle sensor. After reviewing the current state-of-the art calibration procedure, we develop calibration strategies both for the case of a non-harmonic trapping potential, and for the situation where the effective bath temperature is not known. Finally, in Sec.~\ref{sec:energy-measurement}, we describe how to properly measure the center-of-mass temperature of an optically trapped nanoparticle. We focus in particular on the measurement uncertainty that is caused by the thermal fluctuations of the energy, and discuss the repercussions of inevitable drifts in the experimental system.

\section{Experimental setup}\label{sec:Exp_setup}
We experimentally demonstrate all suggested calibration procedures using the system shown in Fig.~\ref{fig:setup}(a). A spherical silica nanoparticle (nominal radius $r=68\,\mathrm{nm}$) is trapped in a strongly focused laser beam (wavelength $\lambda=1064\,\mathrm{nm}$, power $P=80\,\mathrm{mW}$).\cite{Gieseler2012} The trap is placed inside a vacuum chamber, which allows us to control the gas damping rate that the levitated particle experiences. We collect the light scattered from the particle with a lens and send it to a balanced photo detector to measure the particle displacement. This detected signal $V$ is directly proportional to the particle's displacement $q$ for particle displacements typically encountered when the oscillator is at room temperature. We note that the center-of-mass motion of a trapped particle has three degrees of freedom. Since our discussion is valid for any degree of freedom, it is formulated for a single degree of freedom for clarity.

\section{Calibration of detector signal}\label{sec:calibration-of-the-detected-signal}
Any strategy to calibrate the measured displacement signal relies on measuring the oscillator's response to a known force. For calibration, we consider two different types of forces. First, there are fluctuating forces stemming from the coupling to a thermal bath. \rev{Although random}, these forces fulfill certain statistical properties that can be harnessed for calibration, as discussed in Sec.~\ref{sec:fluctuating-forces}. The second type of \rev{forces} we consider are harmonic driving forces with well-defined amplitude, frequency, and phase. We harness such single-tone forces for calibration in Sec.~\ref{sec:harmonic-force}.

Let us start our discussion by considering the classical equation of motion for the position $q$ of a thermally and harmonically driven, damped oscillator
\begin{equation}\label{eq:equation_of_motion}
m\ddot{q}+m\gamma\dot{q}=F_\mathrm{re}(q)+F_\mathrm{dr}(t)+F_\mathrm{fluct}(t),
\end{equation}
where $\gamma$ is the damping rate, $F_\mathrm{re}$ is a general restoring force, $F_\mathrm{dr}$ is a harmonic driving force, $m$ is the mass of the oscillator, and $\dot{q}$ and $\ddot{q}$ denote the first and second time derivatives of $q$, respectively.
%\cite{Pinard1999}
%For optically levitated spherical particles in vacuum, the effective mass can be deduced from the measured damping rate using kinetic gas theory.\cite{Beresnev1990,Gieseler2012}
The fluctuating force $F_\mathrm{fluct}$ fulfills the fluctuation-dissipation theorem $\langle F_\mathrm{fluct}(t)F_\mathrm{fluct}(t+\tau)\rangle=2m\gamma k_\mathrm{B}T_\mathrm{b}\delta(\tau)$ with the Boltzmann constant $k_\mathrm{B}$, and the Dirac distribution $\delta$.\cite{Langevin1908} In thermal equilibrium, \rev{the fluctuating force and the damping balance out, leading} to a steady state of the oscillator motion characterized by the bath temperature $T_\mathrm{b}$. This fact can be used for calibration given that the system is in equilibrium with a bath of known temperature.\cite{Hutter1993,Hauer2013}

\subsection{Calibration using fluctuating forces}\label{sec:fluctuating-forces}
We first turn our attention to reviewing the state-of-the-art calibration technique that relies on the fluctuating forces stemming from a thermal bath acting on a {harmonic} oscillator in Sec.~\ref{sec:harmonic-oscillator}, before providing a calibration strategy for anharmonic potentials in Sec.~\ref{sec:nonlinear-oscillator}.

\subsubsection{Calibration for harmonic trapping potential}\label{sec:harmonic-oscillator}
For a harmonic oscillator, the restoring force is given by $F_\mathrm{re}(q)=-k q$ with the spring constant $k=m\Omega_0^2$ and the natural oscillation frequency $\Omega_0$.
Using the equation of motion Eq.~(\ref{eq:equation_of_motion}) with $F_\mathrm{dr}=0$, we find the single-sided power spectral density for a harmonic oscillator\rev{\cite{Li2011,Kiesel2013,Hauer2013,Li2013,Gieseler2014b,Millen2014,Frangeskou2016,Fonseca2016,Rondin2017}}
\begin{equation}\label{eq:PSD_harmonic}
\hat{S}_{qq}(\Omega)=\frac{2k_\mathrm{B}T_\mathrm{b}\gamma/(\pi m)}{(\Omega_0^2-\Omega^2)^2+\gamma^2\Omega^2}.
\end{equation}
According to the Wiener-Khinchin theorem, the variance is related to the power spectral density by the relation
\begin{equation}\label{eq:var_PSD}
\langle q^2\rangle = \int_{0}^{\infty}\d\Omega\, \hat{S}_{qq}(\Omega).
\end{equation}
Using Eq.~(\ref{eq:PSD_harmonic}), we obtain $m\Omega_0^2\langle q^2\rangle = k_\mathrm{B}T_\mathrm{b}$.
This result is equivalent to the equipartition theorem, which states that the mean potential energy of every oscillation mode of a harmonic oscillator in thermal equilibrium is given by
\begin{equation}\label{eq:equipartition_Epot}
\langle E_\mathrm{pot}\rangle=\frac{1}{2}m\Omega_0^2\langle q^2\rangle=\frac{1}{2}m\Omega_0^2\frac{\langle V^2\rangle}{c_\mathrm{calib}^2}=\frac{1}{2}k_\mathrm{B}T_\mathrm{b}.
\end{equation}
Accordingly, if the bath temperature $T_\mathrm{b}$, the mass $m$, and the natural frequency $\Omega_0$ are known, a measurement of $\langle V^2\rangle$ allows us to retrieve the calibration factor $c_\mathrm{calib}$.

To summarize, for a \emph{harmonic} oscillator in thermal equilibrium, the calibration factor may be determined in two equivalent ways:
\begin{enumerate}
	\item Calculate the variance of the detector signal time trace $V(t)$ directly, or integrate over the power spectral density $\hat{S}_{VV}(\Omega)$ according to Eq.~(\ref{eq:var_PSD}). It can be beneficial to spectrally filter the signal by integrating the power spectral density over a limited frequency band to exclude technical noise outside that band. Then, calculate the calibration factor $c_\mathrm{calib}$ from Eq.~(\ref{eq:equipartition_Epot}) using the mass $m$, the bath temperature $T_\mathrm{b}$, and the oscillation frequency $\Omega_0$ that is extracted from the power spectral density.
	\item Calculate the power spectral density of the detected signal $\hat{S}_{VV}(\Omega)=c_\mathrm{calib}^2\hat{S}_{qq}(\Omega)$, fit Eq.~(\ref{eq:PSD_harmonic}), and deduce the calibration factor $c_\mathrm{calib}$ using the mass $m$ and the bath temperature $T_\mathrm{b}$.
\end{enumerate}

\begin{figure}
	\includegraphics{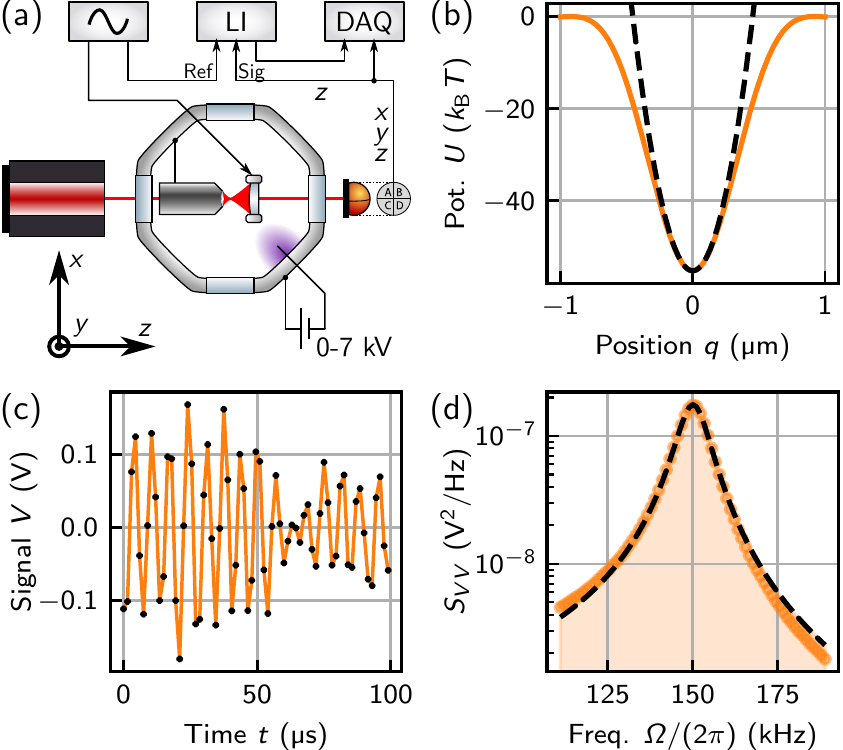}
	\caption{\emph{Optically levitated oscillator system:} \textbf{(a)}~Experimental setup for optical trapping of a nanoparticle in vacuum. A laser is focused with an objective to form the optical trap for the particle. The light scattered by the particle is collected with a lens and detected with a balanced detector. The detector signal is recorded with a data acquisition card (DAQ). \textbf{(b)}~Calculated potential along the $y$ axis for a strongly focused beam with focal power $80\,\mathrm{mW}$ and numerical aperture $\mathrm{NA}=0.8$ (\rev{solid orange line}). The harmonic component of the potential is shown as the \rev{dashed black} line. \textbf{(c)}~Part of the recorded detector signal time trace representing the thermally driven particle motion along the $y$ direction at a pressure of $10(1)\,\mathrm{mbar}$ (black dots). \textbf{(d)}~The power spectral density of the detector signal. The \rev{dashed black} line is a fit of Eq.~(\ref{eq:PSD_harmonic}) to the data. \label{fig:setup}}
\end{figure}

As an illustration, we now experimentally implement this calibration procedure based on fluctuating forces in the harmonic-oscillator approximation, using our experimental setup sketched in Fig.~\ref{fig:setup}(a). We plot the calculated optical trapping potential $U(q)\propto{-}I(q)$ in Fig.~\ref{fig:setup}(b), where $I(q)$ is the optical intensity of a strongly focused laser beam (\rev{solid orange} line).\cite{Novotny2012} For comparison, we also plot the parabolic component of the potential (\rev{dashed black line}). We experimentally record a time trace $V(t)$ of the particle's oscillation in the optical potential at a gas pressure of $10(1)\,\mathrm{mbar}$ [a short section is shown in Fig.~\ref{fig:setup}(c)] and derive the power spectral density of the detector signal, plotted in Fig.~\ref{fig:setup}(d). To determine the calibration factor, we fit Eq.~(\ref{eq:PSD_harmonic}) to the power spectral density. For a sphere in a viscous medium, we can deduce the sphere radius from the damping rate $\gamma$ retrieved from the fit.\cite{Beresnev1990,Gieseler2012} Together with the mass density of silica $\rho=2200\,\mathrm{kg/m^3}$, we extract the particle mass $m=1.6(5)\,\mathrm{fg}$ and obtain the calibration factor $c_\mathrm{calib}=0.95(15)\,\mathrm{mV/nm}$.

\subsubsection{Calibration for anharmonic trapping potential}\label{sec:nonlinear-oscillator}
Having discussed the calibration method using the \emph{harmonic} oscillator approximation, \rev{we provide in this section} a calibration procedure for an \emph{anharmonic} trapping potential, i.e., \rev{in} the case of a general restoring force $F_\mathrm{re}(q)$. For an anharmonic potential, equipartition of the \emph{potential} energy in Eq.~(\ref{eq:equipartition_Epot}) does not generally hold. In particular, if the anharmonic potential couples different oscillation modes, we cannot even assign a potential energy to a single mode anymore. Therefore, in general, the calibration technique described in Sec.~\ref{sec:harmonic-oscillator} using the potential energy is not valid for the anharmonic potentials that every realistic oscillator system exhibits. Nevertheless, we can calibrate the detected signal using the \emph{kinetic} energy of the oscillator, for which equipartition amongst the different degrees of freedom still holds in the form
\begin{equation}\label{eq:equipartition_Ekin}
\langle E_\mathrm{kin}\rangle=\frac{1}{2}m\langle \dot{q}^2\rangle=\frac{1}{2}m\frac{\langle \dot{V}^2\rangle}{c_\mathrm{calib}^2}=\frac{1}{2}k_\mathrm{B}T_\mathrm{b}
\end{equation}
regardless of anharmonicities or coupling between modes.

To derive $\langle \dot{V}^2\rangle$, \rev{one usually cannot} rely on a direct measurement of $\dot{V}$, which is proportional to the velocity of the particle. However, the variance $\langle \dot{V}^2\rangle$ can be conveniently calculated by numerically integrating the power spectral density $\hat{S}_{\dot{V}\dot{V}}$ in analogy to Eq.~(\ref{eq:var_PSD}). The power spectrum $\hat{S}_{\dot{V}\dot{V}}$ can be obtained (even in the case of an anharmonic oscillator with an arbitrary potential) from the displacement power spectral density as\cite{Schliesser2008,Faust2012}
\begin{equation}\label{eq:velocity_PSD}
\hat{S}_{\dot{V}\dot{V}}(\Omega) = \Omega^2 \hat{S}_{VV}(\Omega).
\end{equation}
In practice, we have to consider that at high frequencies the integration of technical measurement noise, which is then also \rev{multiplied by} $\Omega^2$, can result in overestimating the variance of the velocity. This effect can be reduced by limiting the integration band to exclude high frequency noise.

In the example of Fig.~\ref{fig:setup}(d), we find the displacement calibration factor $c_\mathrm{calib}=0.92(15)\,\mathrm{mV/nm}$ using the approach via the kinetic energy in Eqs.~(\ref{eq:equipartition_Ekin}) and~(\ref{eq:velocity_PSD}). This calibration factor is $3\%$ smaller than the one derived in Sec.~\ref{sec:harmonic-oscillator}, where we assumed our trapping potential to be strictly harmonic. This means that using the harmonic oscillator approximation from Sec.~\ref{sec:harmonic-oscillator} results in the energy being underestimated by $6\%$ (cf. Sec.~\ref{sec:energy-measurement}).
Naturally, it depends on the required measurement precision whether this calibration error is permissible. However, we stress that the calibration strategy using Eqs.~(\ref{eq:equipartition_Ekin}) and~(\ref{eq:velocity_PSD}) is always preferred over that detailed in Sec.~\ref{sec:harmonic-oscillator}, since it is correct both in presence and absence of anharmonicities in the trapping potential while requiring no additional measurement effort.

%To assess the contribution of nonlinearities, we consider Duffing nonlinearities, i.e., $q^4$ terms in the oscillator potential, that are usually the lowest order nonlinearities in mechanical oscillators. Then the restoring force for a specific mode $j$ is $F_\mathrm{re}(t)=-kq_j(t)[1+\sum_i\xi_{ij}q_i^2(t)]$, where $\xi_{ij}$ are the Duffing parameters and we sum over all the modes $i$ (including $i=j$) that are coupled to the oscillation mode of interest.\cite{Lifshitz2008} In order to make an estimate of the influence of the nonlinear part of the force, we consider the time average $\sum_i\xi_{ij}\langle q_i^2\rangle\approx \sum_iE_i\xi_{ij}/(m \Omega_\mathrm{0i}^2)$ with the oscillation frequency $\Omega_\mathrm{0i}$ of the mode $i$. Whenever this expression becomes significant compared to $1$, the contribution of nonlinearities to the oscillator motion are important and it is necessary to use the kinetic energy for calibration, as discussed in this section. The above estimation also implies that the impact of the nonlinearities depends on the temperature of all coupled modes and eventually becomes negligible when the oscillator is cooled. In our example Fig.~\ref{fig:setup}, this term is $\sum_i\xi_{ij}\langle q_i^2\rangle=0.04$.

\subsection{Calibration using a harmonic driving force}\label{sec:harmonic-force}
So far, we considered a spectrally white thermal force acting on the oscillator arising \rev{from} a coupling to a bath of known temperature.
Unfortunately, however, this effective bath temperature is frequently not known. In particular, levitated nanoparticles have been shown to acquire a considerable internal temperature at reduced pressure, where convective cooling by the surrounding bath ceases to be efficient.\cite{Millen2014} The internally hot particle heats the residual gas around it, creating an effective thermal bath at a temperature different from that of the vacuum chamber. This means that at the reduced pressures where levitated optomechanical sensors typically operate a calibration against a thermal bath is not feasible. In this section, we eliminate this problem by performing a calibration using a harmonic driving force.\rev{\cite{Bushev2013,Ranjit2015}} In contrast to a calibration procedure recently demonstrated for levitated microspheres, where a harmonic driving force is applied \rev{far below} resonance,\cite{Rider2017} we operate close to the mechanical resonance of the levitated nanoparticle.
We note that this calibration method requires the trapping potential to be harmonic, at least in the region sampled by the particle during its motion. Fortunately, feedback cooling of optically levitated nanoparticles is \rev{well-established} as a means to reduce the effective temperature of the particle's center-of-mass motion to a regime where the particle does not sample the anharmonicities of the potential.\cite{Li2011,Gieseler2012,Kiesel2013,Millen2015,Vovrosh2017} Under suitable feedback cooling, the trapped particle can be viewed as a harmonic oscillator whose damping rate is dominated by the feedback.\cite{Jain2016} Accordingly, for strongly anharmonic potentials, the calibration procedure discussed in this section can be applied to \rev{a} particle under feedback.

Our calibration strategy relies on \rev{applying} a harmonic driving force $F_\mathrm{dr}(t)=F_0\sin(\Omega_\mathrm{dr}t)$ at \rev{a} frequency $\Omega_\mathrm{dr}$ to the oscillator governed by the equation of motion~Eq.~(\ref{eq:equation_of_motion}). \rev{Assuming a} restoring force $F_\mathrm{re}(q)=-kq$ and using a harmonic ansatz for Eq.~(\ref{eq:equation_of_motion}), the variance of the oscillator's displacement in response to the force $F_\mathrm{dr}$ is
\begin{equation}\label{eq:amplitude_extForce}
\langle q^2\rangle = \frac{\langle V^2\rangle}{c_\mathrm{calib}^2} = \frac{F_0^2/(2m^2)}{(\Omega_0^2-\Omega_\mathrm{dr}^2)^2+\gamma^2\Omega_\mathrm{dr}^2}.
\end{equation}
Accordingly, we can extract the calibration factor $c_\mathrm{calib}$ using the mass $m$ and the amplitude of the force $F_0$, \rev{requiring no prior} knowledge of the effective bath temperature $T_\mathrm{b}$. The center frequency $\Omega_0$ and the damping rate $\gamma$ are extracted from a fit of Eq.~(\ref{eq:PSD_harmonic}) to the power spectral density.

\subsubsection*{Example: Coulomb force on levitated nanoparticle}
To implement a harmonic driving force, and to demonstrate the corresponding calibration protocol, we make use of the fact that the levitated particle can be controllably charged.\cite{Moore2014,Frimmer2017} As sketched in Fig.~\ref{fig:setup}(a), we use the objective and the holder of the collection lens to form a capacitor around the particle, \rev{and apply} a Coulomb force of the type $F_\mathrm{dr}(t)=Q\mathcal{E}_0\sin(\Omega_\mathrm{dr}t)$ to our oscillator, where $Q$ is the particle's charge, and $\mathcal{E}_0$ is the electric field along the oscillation direction.\cite{Ranjit2015} We prepare a charge of $|Q|=10\, Q_\mathrm{e}$ on the particle, where $Q_\mathrm{e}$ is the elementary charge, and apply an electric field along the $z$ direction with amplitude $\mathcal{E}_0=360\,\mathrm{V/m}$. The simultaneous use of three drive tones at 25, 45 and 50~kHz allows us to increase our measurement precision by averaging.

\begin{figure}
	\includegraphics{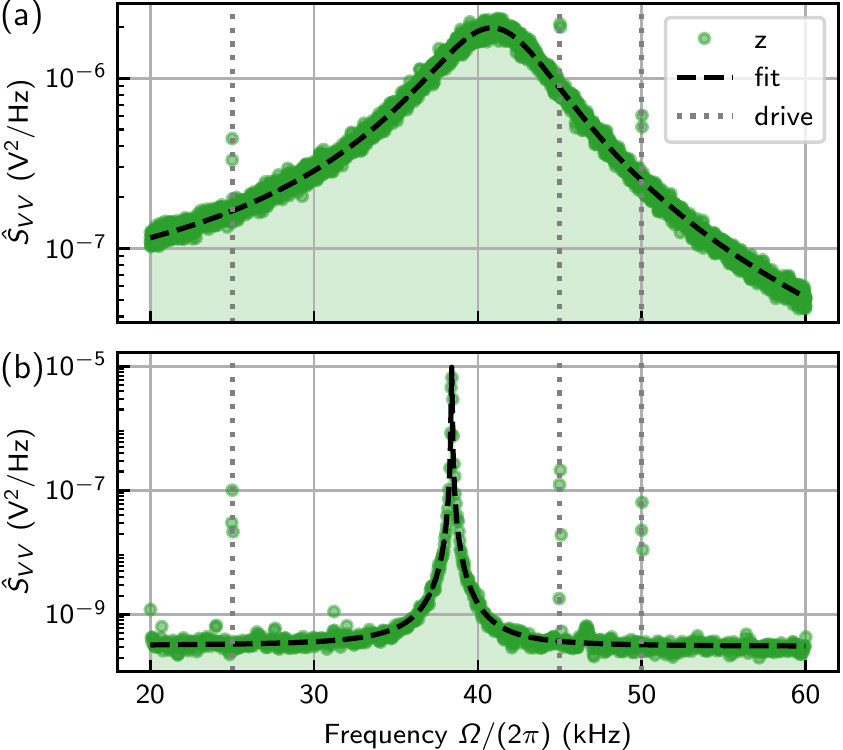}
	\caption{\emph{Calibration using a harmonic Coulomb force:} \textbf{(a)}~Power spectral density (PSD) at $10(1)\,\mathrm{mbar}$ showing the thermally driven resonance of the uncooled oscillator [fit of Eq.~(\ref{eq:PSD_harmonic}) \rev{as dashed black line}] with the response to the monochromatic driving force (\rev{dotted gray lines}). \textbf{(b)}~PSD of the feedback-cooled oscillator at a pressure of $1.0(2)\times10^{-3}\,\mathrm{mbar}$. \label{fig:charge_calibration}}
\end{figure}

We apply \rev{this calibration procedure} at pressures below $10^{-3}\,\mathrm{mbar}$ and under feedback cooling, where a calibration against the thermal fluctuating force is not possible \rev{due to a lack of knowledge of} the effective bath temperature. Before moving to lower pressures, as a cross-check, we first benchmark the \rev{harmonic-driving-forces calibration} against the \rev{fluctuating-forces calibration} from Sec.~\ref{sec:fluctuating-forces}. To do so, at a pressure of $10(1)\,\mathrm{mbar}$, we record a \rev{$15\,\mathrm{s}$ time trace of the detector signal $V(t)$} and plot the power spectral density $\hat{S}_{VV}(\Omega)$ in Fig.~\ref{fig:charge_calibration}(a). From a fit of Eq.~(\ref{eq:PSD_harmonic}) (\rev{dashed black line}) we find the center frequency $\Omega_0=2\pi\times 41\,\mathrm{kHz}$ and the linewidth $\gamma=2\pi\times 7.7\,\mathrm{kHz}$.
%We note that using the harmonic approximation introduces a systematic error, as the optical potential is slightly nonlinear at room temperature. %(although here the harmonic oscillator approximation is not strictly valid as discussed before).
From the linewidth, we derive the particle mass $m=3.2(10)\,\mathrm{fg}$ using the gas law, as explained before.\cite{Gieseler2012,Beresnev1990} The presence of the harmonic Coulomb force at three drive frequencies acting on the particle gives rise to the sharp signatures in the power spectrum marked with \rev{dotted gray} lines in Fig.~\ref{fig:charge_calibration}. We integrate the detector signal in a narrow band of $20\,\mathrm{Hz}$ around the driving frequencies and obtain a signal variance $\langle V^2\rangle$ at each of the \rev{three frequencies}. %$\langle v^2\rangle^{(i)} \in \{11,50,14\}\,\mathrm{mV^2}$.
Using Eq.~(\ref{eq:amplitude_extForce}), we finally extract a calibration factor of $c_\mathrm{calib}=1.05(32)\,\mathrm{mV/nm}$ (averaged over the three drive tones).
%(The error in $c_\mathrm{calib}$ mainly originates from the uncertainty of the effective mass $m$.) An additional error arises ...

This result is in good agreement with the calibration using fluctuating forces and the kinetic energy proportional to $\langle \dot{V}^2\rangle$ following Sec.~\ref{sec:nonlinear-oscillator}, which yields $c_\mathrm{calib}=1.07(17)\,\mathrm{mV/nm}$ for the power spectral density in Fig.~\ref{fig:charge_calibration}(a). The small discrepancy can be explained by the fact that for this cross-check we had to assume a harmonic trapping potential for the calibration using a Coulomb force, an assumption that is not strictly valid for our experimental system at room temperature, as was shown in Secs.~\ref{sec:harmonic-oscillator} and~\ref{sec:nonlinear-oscillator}.

Having established that the calibration obtained using a harmonic driving force is in agreement with our previous calibrations, we now reduce the pressure to $1.0(2)\times10^{-3}\,\mathrm{mbar}$ and activate parametric feedback cooling to reduce the oscillation amplitude.\cite{Gieseler2012} Accordingly, the effective bath temperature $T_\mathrm{b}$, which is reduced due to the feedback, is not known and a calibration against the fluctuating forces is therefore not possible. Under these conditions the influence of the nonlinearities is small enough to be neglected and we can safely assume a harmonic oscillator. To calibrate using the harmonic Coulomb force, we record a \rev{$15\,\mathrm{s}$ time trace} using the same harmonic driving force at three tones as before. The measured power spectral density is plotted in Fig.~\ref{fig:charge_calibration}(b). As the oscillator linewidth at this pressure with activated feedback is only $\gamma\approx2\pi\times 20\,\mathrm{Hz}$ and the center frequency is drifting on the same scale due to fluctuations in the trapping laser power, pointing direction, and polarization, we split the long time trace into $20$ sections and determine a calibration factor $c_\mathrm{calib}$ for each section, following the procedure established at high pressure. Averaging the calibration factors for all sections and the three driving frequencies yields $c_\mathrm{calib}=0.66(20)\,\mathrm{mV/nm}$. This calibration factor is a factor $1.6$ smaller than the calibration factor at high pressure. This is an important finding. It means that in our experimental system the calibration factor, relating the detector signal to the particle's displacement, changes when the pressure in the vacuum chamber is reduced. This observation underlines the need to calibrate any sensor under the final operating condition. Currently, we can only speculate that this change in the calibration factor is related to a change of the internal particle temperature modifying the particle's optical properties,\cite{Millen2014} or a deformation of the optical components when reducing the gas pressure. \rev{To confirm this conjecture, independent control of both the internal particle temperature and the temperature of the optical elements would be necessary.}

\section{Mode-temperature measurement}\label{sec:energy-measurement}
In numerous deployments, a levitated-nanoparticle sensor is used to probe a thermal state, often while working towards the goal of bringing a levitated nanoparticle into the quantum ground state of motion.\cite{Gieseler2012,Kiesel2013,Millen2015,Jain2016,Vovrosh2017} For any experiment involving the cooling of the levitated particle's center-of-mass motion, an accurate and precise measurement of the temperature characterizing the particle's thermal state is essential. This temperature measurement is challenging for two reasons. For a harmonic oscillator the mode temperature $T$ is linked to the mean total oscillation energy $\langle E\rangle$ in the form $T=\langle E\rangle/k_\mathrm{B}$. However, in the case of an anharmonic oscillator, this simple relation does not hold anymore. In addition, for an oscillator in a thermal state, we are confronted with an oscillator energy that fluctuates over time. Therefore, it is not straight forward to determine a temperature from a position measurement.

In this section, we first discuss the calculation of the mode temperature from a recorded time trace $V(t)$. We then turn our attention to the uncertainty in the mode-temperature measurement. As the energy is a fluctuating quantity, it does not only exhibit statistical and systematic errors, but also uncertainties that originate from the limited duration for which we observe the thermal state. Therefore, we discuss in particular the influence of the measurement time on the precision of the mode-temperature measurement.

\subsection{Mode temperature of anharmonic oscillator}

We are interested in calculating the temperature $T$ of the oscillation mode from the measured displacement time trace $V(t)$. If the potential is harmonic, we can express the mode temperature using the potential energy
\begin{equation}\label{eq:E_pot}
T = \frac{2\langle E_\mathrm{pot}\rangle}{k_\mathrm{B}}=m\Omega_0^2\frac{\langle V^2\rangle}{k_\mathrm{B}c_\mathrm{calib}^2}.
\end{equation}

However, as discussed in Sec.~\ref{sec:nonlinear-oscillator}, for anharmonic potentials, Eq.~(\ref{eq:E_pot}) is not generally valid anymore and we have to use the kinetic energy for deriving the mode temperature
\begin{equation}\label{eq:E_kin}
T = \frac{2\langle E_\mathrm{kin}\rangle}{k_\mathrm{B}}=m\frac{\langle \dot{V}^2\rangle}{k_\mathrm{B}c_\mathrm{calib}^2}.
\end{equation}
For deriving $\langle \dot{V}^2\rangle$, we can again integrate the velocity power spectral density $\hat{S}_{\dot{V}\dot{V}}$ following Eq.~(\ref{eq:velocity_PSD}).

The systematic error of the temperature measurement is given by the errors in the mass $m$ and the calibration factor $c_\mathrm{calib}$. Importantly, the calibration against a fluctuating force using Eqs.~(\ref{eq:equipartition_Epot}) and~(\ref{eq:equipartition_Ekin}) allows us to define an energy calibration factor $C_\mathrm{calib}=c_\mathrm{calib}^2/m$, such that $T = \langle \dot{V}^2\rangle/(k_\mathrm{B}C_\mathrm{calib})$, providing the advantage that the knowledge of the mass is not required. In contrast, the calibration using a harmonic driving force and a subsequent mode-temperature measurement, always require the knowledge of the mass and any uncertainty of the mass reflects in the uncertainty of the measured energy.

\rev{We again stress the advantage} of using the kinetic energy in thermal equilibrium for calibration as detailed in Sec.~\ref{sec:nonlinear-oscillator}.
When using the potential energy Eq.~(\ref{eq:E_pot}) for calibration as reviewed in Sec.~\ref{sec:harmonic-oscillator}, any error in the measurement of the frequency $\Omega_0$ enters the uncertainty of the mode temperature. This error can be significant in the presence of anharmonicities of the trapping potential, which leads to an amplitude dependent shift of the oscillation frequency.\cite{Lifshitz2008}

\subsection{Temperature uncertainty due to thermal fluctuations}\label{sec:measurement-time}

The uncertainty associated with a measurement of the mode temperature is not only set by the error in the calibration factor and the mass, but also by the length $\tau$ of the measured time trace $V(t)$. This is the case because in a thermal state the energy is a fluctuating quantity due to the fluctuating forces that act on the oscillator. Therefore, even for an ideal measurement in the absence of measurement noise, we can regard the variance $\overline{V^2}$ of a time trace $V(t)$ of finite length $\tau$ only as an estimate for the corresponding expectation value $\langle V^2\rangle$. In this section, we answer the following question: How long do we need to observe the oscillator's motion in order to determine the mode temperature $T$ with a particular relative standard deviation $\sigma_{T}/T$? In our treatment, we first focus exclusively on the fundamental measurement uncertainties arising from the fluctuations of the energy in a thermal state before turning to limitations imposed by experimental difficulties.

We start by providing a theoretical treatment of the uncertainty of a temperature measurement. In general, for any mechanical oscillator, we can derive the relative standard deviation of the mode temperature from stochastic arguments, considering that the energy of a thermal state is Boltzmann distributed. If we calculate the mode temperature from a time trace of length $\tau$ at a damping rate $\gamma$, the relative standard deviation is given by\cite{Frenkel2001}
\begin{equation}\label{eq:energy_uncertainty}
\frac{\sigma_{T}}{T}=\frac{\sigma_{\overline{V^2}}}{\overline{V^2}}=\sqrt{\frac{2}{\gamma\tau}}.
\end{equation}
This equation takes into account that consecutive measurements are correlated over a time $1/\gamma$, leading to an effectively smaller sample size. From Eq.~(\ref{eq:energy_uncertainty}), the measurement time $\tau$ that is required to determine the mode temperature with a desired error $\sigma_{T}/T$ at a damping rate $\gamma$ can be calculated.
The relative standard deviation only contains the product ${\gamma\tau}$, which illustrates that the time scale of the energy fluctuations is set by the damping rate $\gamma$. As an example, to reach a relative standard deviation for the mode-temperature measurement of below 1\% at a damping rate of $\gamma=2\pi\times 1\,\mathrm{kHz}$, a time trace of at least \rev{$3\,\mathrm{s}$ is required}.

\begin{figure}
	\includegraphics{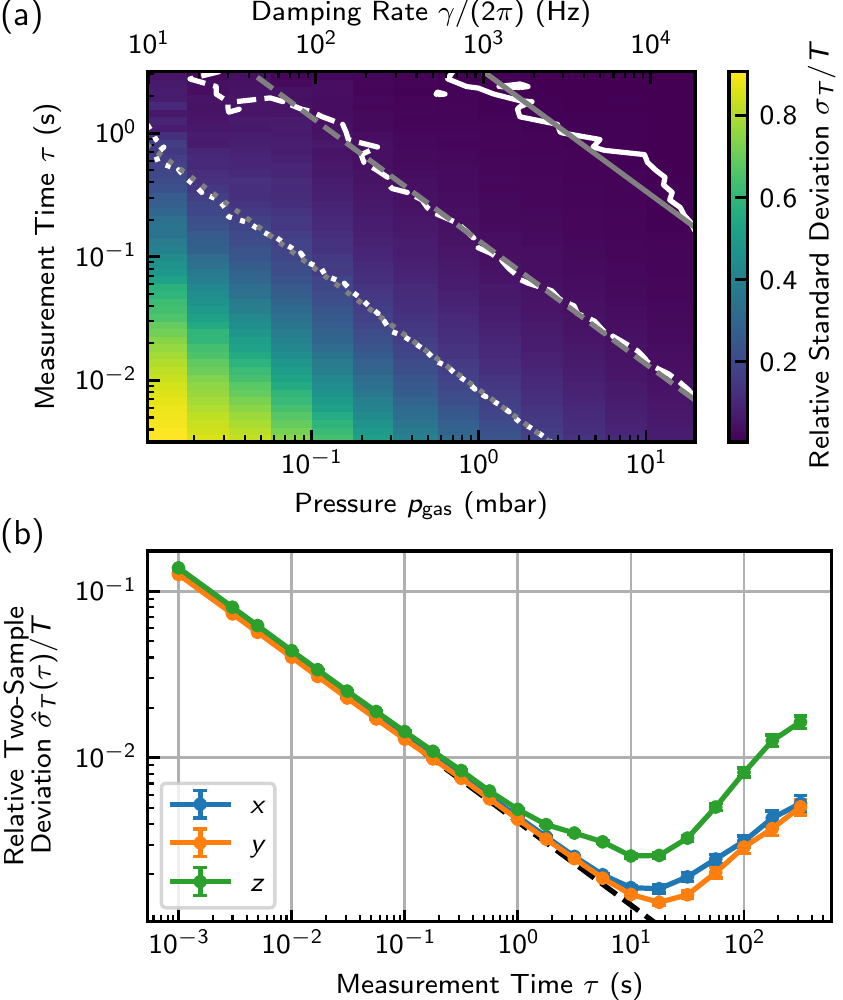}
	\caption{\emph{Uncertainty of the mode-temperature measurement:} \textbf{(a)}~Relative standard deviation of the center-of-mass temperature of a levitated particle derived from measurements of duration $\tau$ of the particle displacement using Eq.~(\ref{eq:E_pot}).  This standard deviation reduces for increasing gas pressure $p_\mathrm{gas}$ and measurement time $\tau$. Marked in white are the measured temperature uncertainties of $1\,\%$ (solid), $5\,\%$ (dashed), and $20\,\%$ (dotted). For comparison, calculated uncertainties according to Eq.~(\ref{eq:energy_uncertainty}) are plotted in gray (no fit). \textbf{(b)}~The two-sample deviation of the mode-temperature measurement calculated from a $4\,\mathrm{h}$ time trace. For short measurement times the two-sample deviation follows Eq.~(\ref{eq:energy_uncertainty}) (\rev{dashed black line}). For measurement times longer than $1\,\mathrm{s}$, $\hat{\sigma}_{T}(\tau)$ deviates from the $\tau^{-1/2}$ trend, which indicates drifts in the experimental setup. \label{fig:measurement_time}}
\end{figure}

After theoretically answering the question of how long we have to measure to achieve a given temperature uncertainty, we experimentally determine the relative standard deviation of the mode-temperature measurement in our levitated particle system. To this end, we record $N$ time traces of length $\tau$, calculate the estimate for the mode temperature from each time trace, and then derive the standard deviation of these temperature measurements. For practical reasons, we record a single \rev{$18\,\mathrm{s}$-long time trace} and split it into $N$ blocks of length $\tau$. For each of these blocks, we compute the mode temperature $T_i$ according to Eq.~(\ref{eq:E_kin}). The standard deviation $\sigma_{T}$ of $N$ measurements of the mode temperature $T_i$, normalized by their mean value $T=\sum_i T_i/N$, then gives the relative standard deviation of the mode-temperature. In Fig.~\ref{fig:measurement_time}(a), $\sigma_{T}/T$ is plotted for different values of the measurement time $\tau$ and gas pressure $p_\mathrm{gas}$, which is proportional to the damping rate $\gamma$. We find that the measurement uncertainty decreases when either the measurement time $\tau$ or the damping rate $\gamma$ is increased. In Fig.~\ref{fig:measurement_time}(a), we mark with white lines the experimentally measured mode temperature uncertainty of $1\,\%$ (solid), $5\,\%$ (dashed), and $20\,\%$ (dotted). For comparison, we plot in gray the expected relative standard deviations calculated from Eq.~(\ref{eq:energy_uncertainty}), which are in good agreement with the measured results.

\subsection{Practical limitations and optimal measurement time}

According to our considerations in Sec.~\ref{sec:measurement-time}, \rev{longer observation of a thermal state} will always lead to a more precise measurement of its temperature. In practice, however, the temperature measurement is also affected by drifts in the experimental apparatus. These include drifts that impact the oscillator directly and drifts in the measurement system. A common way to quantify drifts in the frequency of an oscillator is the Allan variance, or two-sample variance.\cite{Allan1966} Here, we apply the same concept to the mode-temperature measurement.\rev{\cite{Purdy2017}} We write the two-sample variance of the temperature as
\begin{equation}\label{eq:two-sample_energy}
\hat\sigma_T^2(\tau)=\frac{1}{N-1}\sum_{k=1}^{N-1}\frac{1}{2}\left[T_{k+1}^{(\tau)}-T_{k}^{(\tau)}\right]^2,
\end{equation}
where $T_k^{(\tau)}$ is the estimate of the mode temperature in a section $t\in[k\tau-\tau, k\tau)$ of a long measurement time trace, where $\tau$ is the length of each of the $N$ sections. The two-sample variance for various measurement times $\tau$ can help \rev{identifying} drifts in the experimental setup and reveals the optimal measurement time that yields the smallest measurement uncertainty.

As an example, we record a continuous time trace of the detector signal $V(t)$ for each degree of freedom $(x,y,z)$ of our levitated nanoparticle, with a length of $4\,\mathrm{h}$ at a pressure of $20\,\mathrm{mbar}$ (corresponding to $\gamma=2\pi\times19\,\mathrm{kHz}$). We split this time trace in sections of length $\tau$ and calculate the mode temperature of every section according to Eq.~(\ref{eq:E_kin}). Then, we compute the two-sample variance of the resulting temperature time trace and plot it in Fig.~\ref{fig:measurement_time}(b). For short measurement times, the temperature uncertainty is described by Eq.~(\ref{eq:energy_uncertainty}) (\rev{dashed black} line, no fit), which means that the measurement uncertainty is limited by the thermal fluctuations of the oscillation energy. For measurement times longer than $1\,\mathrm{s}$, the uncertainty starts to deviate from the $\tau^{-1/2}$ trend, which indicates drifts of our experimental apparatus on a time scale of seconds. These drifts have their origin in the stability of our trapping laser's pointing direction, power, and polarization. We find the the optimal duration of a mode-temperature measurement at a measurement time of $10$ to $20\,\mathrm{s}$ for our experimental setup.

\section{Conclusion}
We have discussed different strategies to calibrate the displacement of a levitated-nanoparticle sensor. We conclude that for a calibration using the fluctuating forces arising from a thermal bath, it is essential to apply the equipartition principle to the particle's \emph{kinetic} energy. Importantly, this procedure yields correct results also in the presence of anharmonicities in the trapping potential, in contrast to calibrations invoking equipartition of the potential energy.
Furthermore, we have demonstrated an alternative calibration method using an externally applied harmonic driving force acting on the levitated particle. This method is favorable under conditions where the effective bath temperature is not known. Notably, we found that for our experimental setup the displacement calibration changes when reducing the operating pressure of the sensor. We therefore stress the importance of gauging levitated-nanoparticle sensors \rev{in} their operating conditions for measurements requiring absolute precision. \rev{Finally, we have discussed measurements of the mode temperature of a levitated nanoparticle in thermal equilibrium. The precision of such a measurement is limited by drifts of the measurement apparatus, and by thermal fluctuations of the oscillation energy. For the latter contribution, we provided a rule for the measurement time required to resolve the mode temperature with a given uncertainty.}
Note that our conclusions are not specific to a levitated nanoparticle but apply to any oscillator that is subject to nonlinearities and variable operating conditions.
% If in two-column mode, this environment will change to single-column format so that long equations can be displayed.
%\begin{widetext}
%$$\mbox{put long equation here}$$
%\end{widetext}

% Figures should be put into the text as floats.
% Use the graphics or graphicx packages (distributed with LaTeX2e).
% Put the label that you will use with \ref{} command in the braces of the \label{} command.
%
% \begin{figure}
% \end{figure}

% \begin{table}
% \end{tabular}
% \end{table}

\begin{acknowledgments}
\rev{The authors thank E.~Bonvin for proofreading the manuscript.} This research was supported by the NCCR-QSIT program (no.~51NF40-160591) and the Swiss National Science Foundation (no.~200021L 169319) in cooperation with the Austrian Science Fund (no.~I 3163).
\end{acknowledgments}

\bibliography{references_v2}

\end{document}